\documentclass[aps,prl,twocolumn,superscriptaddress,longbibliography]{revtex4-2}

\usepackage[colorlinks=true,citecolor=blue]{hyperref} 
\usepackage{graphicx}
\usepackage{amsmath}
\usepackage{dcolumn}
\usepackage{bm}

\begin{document}

\title{
Hamiltonian inference from dynamical excitations in confined quantum magnets
}

\author{Netta Karjalainen}
\thanks{These authors contributed equally.}
\affiliation{Department of Chemistry, University of Helsinki, Finland}
\affiliation{Department of Applied Physics, Aalto University, 02150 Espoo, Finland}

\author{Zina Lippo}
\thanks{These authors contributed equally.}
\affiliation{Department of Physics, University of Helsinki, Finland}
\affiliation{Department of Applied Physics, Aalto University, 02150 Espoo, Finland}

\author{Guangze Chen}
\affiliation{Department of Applied Physics, Aalto University, 02150 Espoo, Finland}
\affiliation{Department of Microtechnology and Nanoscience, Chalmers University of Technology, 41296 Gothenburg, Sweden}

\author{Rouven Koch}
\affiliation{Department of Applied Physics, Aalto University, 02150 Espoo, Finland}

\author{Adolfo O. Fumega}
\affiliation{Department of Applied Physics, Aalto University, 02150 Espoo, Finland}

\author{Jose L. Lado}
\affiliation{Department of Applied Physics, Aalto University, 02150 Espoo, Finland}

\date{\today}

\begin{abstract}
Quantum-disordered models provide a versatile platform to explore the emergence of quantum excitations in many-body systems. The engineering of spin models at the atomic scale with scanning tunneling microscopy and the local imaging of excitations with electrically driven spin resonance has risen as a powerful strategy to image spin excitations in finite quantum spin systems. Here, focusing on $S=1/2$ lattices as realized by Ti in MgO, we show that dynamical spin excitations provide a robust strategy to infer the nature of the underlying Hamiltonian. We show that finite-size interference of the dynamical many-body spin excitations of a generalized long-range Heisenberg model allows the underlying spin couplings to be inferred. We show that the spatial distribution of local spin excitations in Ti islands and ladders directly correlates with the underlying ground state in the thermodynamic limit. Using a supervised learning algorithm, we demonstrate that the different parameters of the Hamiltonian can be extracted by providing the spatially and frequency-dependent local excitations that can be directly measured by electrically driven spin resonance with scanning tunneling microscopy. Our results put forward local dynamical excitations in confined quantum spin models as versatile witnesses of the underlying ground state, providing an experimentally robust strategy for Hamiltonian inference in complex real spin models.
\end{abstract}

\maketitle

\section{Introduction}
The engineering of entangled states of matter represents one of the most powerful strategies for designing exotic quantum materials. A paradigmatic example is the case of quantum disordered ground states, generically emerging in quantum spin models,  many-body ground states featuring long-range entanglement and fractional excitations\cite{Savary2016}. Besides a variety of bulk materials hosting potential quantum spin liquid ground states\cite{RevModPhys.88.041002,Banerjee_2017,Broholm_2020,Gao_2019,PhysRevB.100.144432}, an alternative strategy to design these states relies on bottom-up assembling atom by atom\cite{RevModPhys.91.041001,Yang2021}. This strategy has been systematically exploited with scanning tunneling microscopy (STM), with which artificial spin lattices have been assembled with atomic precision \cite{Eigler1990,doi:10.1126/science.1125398,doi:10.1126/science.1214131,doi:10.1126/science.1228519,Khajetoorians2019,Chen2022}. Excitations in these artificial lattices can be measured with inelastic spectroscopy\cite{Heinrich_2004,PhysRevLett.102.256802,Otte2014} and electrically driven paramagnetic resonance\cite{Natterer2017,Yang2021,SciYang2019,Willke_2019,Yang2018,ChenESR2022}. These techniques allow probing the spin excitations locally in space, in contrast with neutron scattering methods used in bulk materials\cite{Banerjee_2017}. However, systematic methodologies for inferring the ground state and microscopic description of a spin system from experimentally available data remain an open problem in quantum materials.

The bottom-up design of quantum spin lattices presents unique opportunities to understand the build-up of quantum disordered states\cite{Savary2016,RevModPhys.91.041001}. In particular, the spatial resolution of STM techniques allows directly inferring the impact of finite size effects on many-body excitations of quantum spin models\cite{Yang2021,RevModPhys.91.041001}. From the experimental point of view, excitations in confined models provide the opportunity of inferring the Hamiltonian of the underlying system by exploiting finite size effects\cite{Crommie_1993,10.21468/SciPostPhys.2.3.020,Khajetoorians_2019}. While this strategy has been widely demonstrated in electronic systems\cite{Crommie_1993,10.21468/SciPostPhys.9.6.085,Huda_2020,Nilius_2002,Gomes_2012,PhysRevResearch.2.043426,Pennec_2007,Yan_2019,Slot_2017,Lobo_Checa_2009}, spin systems represent a much bigger remarkable challenge due to the complex many-body nature of the ground states. Parameters from a Hamiltonian can be extracted in simple cases explicitly\cite{Yang2021}, yet quantum spin models with multiple parameters require more powerful strategies\cite{https://doi.org/10.48550/arxiv.2211.14205}. Machine-learning-powered Hamiltonian learning has arisen as an effective strategy to infer descriptions of complex systems\cite{Wang2017,PhysRevLett.122.020504,Anshu2021,PhysRevResearch.1.033092,PhysRevResearch.4.033223,Gentile2021}. However, their potential for inferring the nature of frustrated quantum spin models from experimentally accessible spatially and frequency-resolved excitations has not been demonstrated.

\begin{figure*}[t!]
\center
\includegraphics[width=\linewidth]{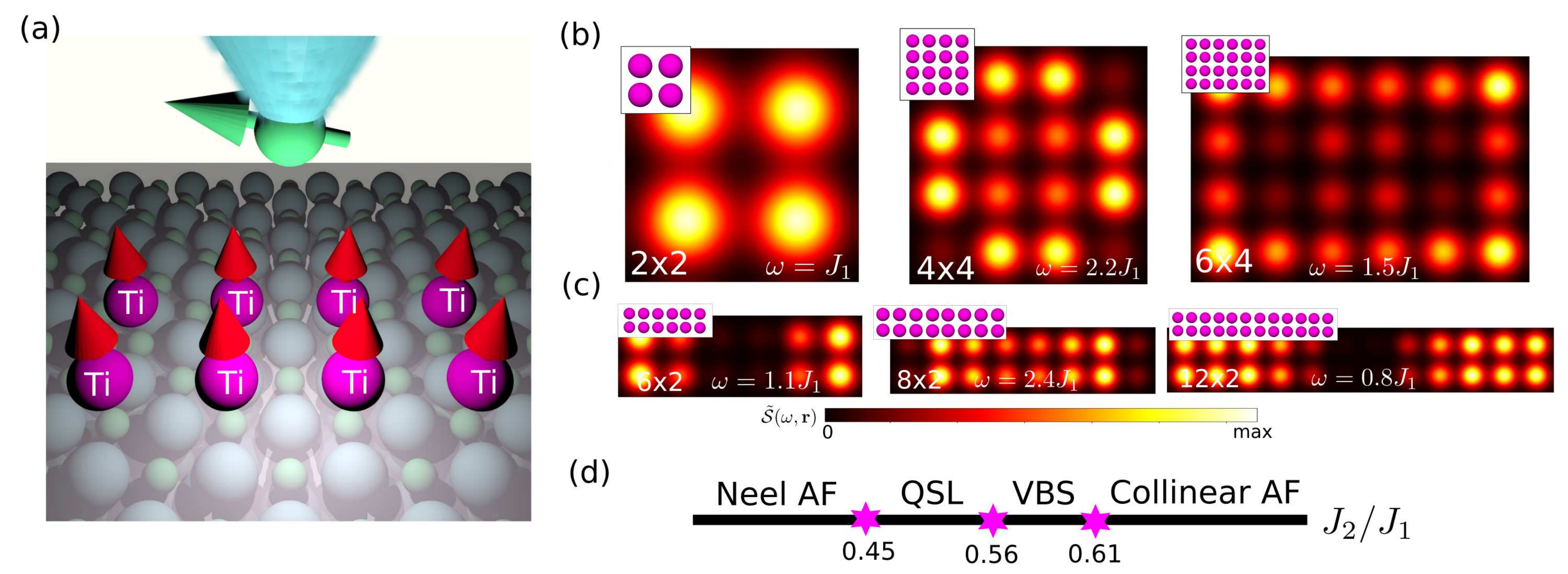}
\caption{(a) Schematic of an artificial spin island of Ti in MgO whose magnetic excitations are measured with ESR-STM. Panels (b,c) show the spatial distribution of spin excitations in different spin islands
at frequencies 
$\omega=J_1$ for $(2\times 2)$,
$\omega=2.2J_1$ for $(4\times 4)$,
$\omega=1.5J_1$ for $(4\times 6)$,
$\omega=1.1J_1$ for $(6\times 2)$,
$\omega=2.4J_1$ for $(8\times 2)$,
$\omega=0.8J_1$ for $(12\times 2)$. 
Panel shows (b) quasi-2D dots of $2\times 2$, $4\times 4$ and $6\times 4$ sizes
and panel (c) and quasi-1D dots of $6\times 2$, $8\times 2$ and $12\times 2$ sizes. Panel (d) shows the magnetic phase diagram realized by the frustrated spin Hamiltonian in the thermodynamic limit of a square lattice (eq. (\ref{eq1})).
}
\label{Fig:schematic}
\end{figure*}

In this manuscript, we show that excitations in finite quantum spin lattices provide a promising strategy to infer the underlying physics of the spin Hamiltonian by exploiting interference from finite-size effects. In particular, we show that a phase transition in the thermodynamic limit has a dramatic impact on the spatial distribution of the many-body modes of finite quantum spin islands, both in the presence and absence of magnetic fields. Furthermore, we demonstrate that this strong dependence allows the development of a strategy to learn the Hamiltonian parameters of spin model directly from the spatially and frequency-dependent excitations, accessible with direct experimental measurements. Our results demonstrate that finite size effects provide a valuable strategy for Hamiltonian learning of quantum spin models.

\section{Magnetic phases in a confined quantum magnet}

In the following, we will focus on the spin
model realized by Ti atoms on MgO\cite{Yang2018,Seifert2020,PhysRevB.104.174408,Yang2021}, recently demonstrated to realize spin excitations in $2\times2$ lattices\cite{Yang2021}.
Figure \ref{Fig:schematic}a shows a schematic of the experimental procedure based on STM and single-atom electron spin resonance (ESR) used to probe these artificial spin islands or quantum spin dots and to measure their spin excitations. 
Different island geometries  with various sizes are illustrated in Figs. \ref{Fig:schematic}b and \ref{Fig:schematic}c, showing the appearance of a spatial pattern in the spin excitations
at specific energies. This pattern in the spin excitation constitutes the fundamental idea for our analysis.

The associated spin Hamiltonian that describes these spin islands takes the form:

\begin{equation} 
\label{eq1}
\mathcal{H}=J_1 \sum_{\langle \mathbf{r}_i,\mathbf{r}_j\rangle}
\mathbf{S}_{\mathbf{r}_i} \cdot \mathbf{S}_{\mathbf{r}_j}
+J_2 \sum_{\langle\langle \mathbf{r}_i,\mathbf{r}_j\rangle\rangle}
\mathbf{S}_{\mathbf{r}_i} \cdot \mathbf{S}_{\mathbf{r}_j},
\end{equation}

where $J_1$ and $J_2$ are the nearest and next-nearest antiferromagnetic spin exchanges, for simplicity, we take $J_1=1$.\footnote{We have considered here the minimal spin Hamiltonian that realizes a frustrated magnet, and that will be our object of study. However, additional terms could be included in this Hamiltonian to study other magnetic phases. For instance, the effect of the Dzyaloshinsky-Moriya interaction has been analyzed within this bottom-up approach\cite{Steinbrecher2018}.} In the thermodynamic limit, this frustrated Hamiltonian develops different magnetic phases depending on the $J_2/J_1$ ratio\cite{PhysRevLett.98.227202,PhysRevLett.93.257206,PhysRevLett.93.177004,PhysRevX.12.031039,PhysRevB.100.125124}, including antiferromagnetic Néel, stripe, quantum spin liquid, and bond ordered, summarized in Fig. \ref{Fig:schematic}d. The antiferromagnetic interactions leading to these phases are the ones that appear in Ti due to superexchange. For Ti in MgO, the couplings $J_1$ and $J_2$ are controlled by the distance between Ti and the thickness of MgO layers.
It is also worth to note that the thickness of MgO may have an impact on the $J_1,J_2,J_3$ parameters due to the contribution to the exchange mediated by the underlying silver substrate in experiments. The impact of silver in the magnetic properties of Ti and other 3d atoms has been addressed previously\cite{PhysRevLett.119.227206,Willke2018,Shehada2022,Tosoni2022}, demonstrating the notable role of the MgO substrate in modifying the effective spin Hamiltonian.

The physical quantity that we will use to characterize the magnetic phase in the confined quantum magnets will be the spin dynamical correlators $\mathcal{S}(\omega,\mathbf{r}_i)$, defined as:

\begin{equation} 
\label{eq2}
    \mathcal{S}(\omega,\mathbf{r}_i)=\langle \text{GS} |
    S_{\mathbf{r}_i}^{z}\delta(\omega+E_{\text{GS}}-\mathcal{H})S_{\mathbf{r}_i}^{z} |\text{GS}\rangle
\end{equation}

where $|\text{GS}\rangle$ and $E_{\text{GS}}$ are the many-body ground state and its energy\cite{ITensor,10.21468/SciPostPhysCodeb.4,dmrgpy}. 
The parameter $\omega$ accounts for the frequency of the excitation, which in ESR experiments corresponds to the frequency of the applied voltage.
It is worth noting that the lineshape of the ESR can be different than the dynamical spin correlator, yet the frequencies at which the features appear will be the same between both quantities.
Eq. \ref{eq2} characterizes local spin excitations, that directly influence inelastic electron tunneling spectroscopy and electron paramagnetic resonance\cite{Andreas2004,Otte2014,Andreas2015,PhysRevLett.119.227206}.
It is worth noting that in both inelastic electron tunneling spectroscopy and electron paramagnetic resonance, additional form factors stemming from transport can slightly modify the signal in comparison with the pure spin spectral function.
\footnote{The spatial profile in the lattice model
is accounted for by a Gaussian envelope of width $r_0$
$
    \tilde{\mathcal{S}}(\omega,\mathbf r)=\sum_i\mathcal{S}(\omega,\mathbf{r}_i)e^{-|\mathbf{r}-\mathbf{r}_i|^2/2r_0}.
$}

\begin{figure}[t!]
\center
\includegraphics[width=\linewidth]{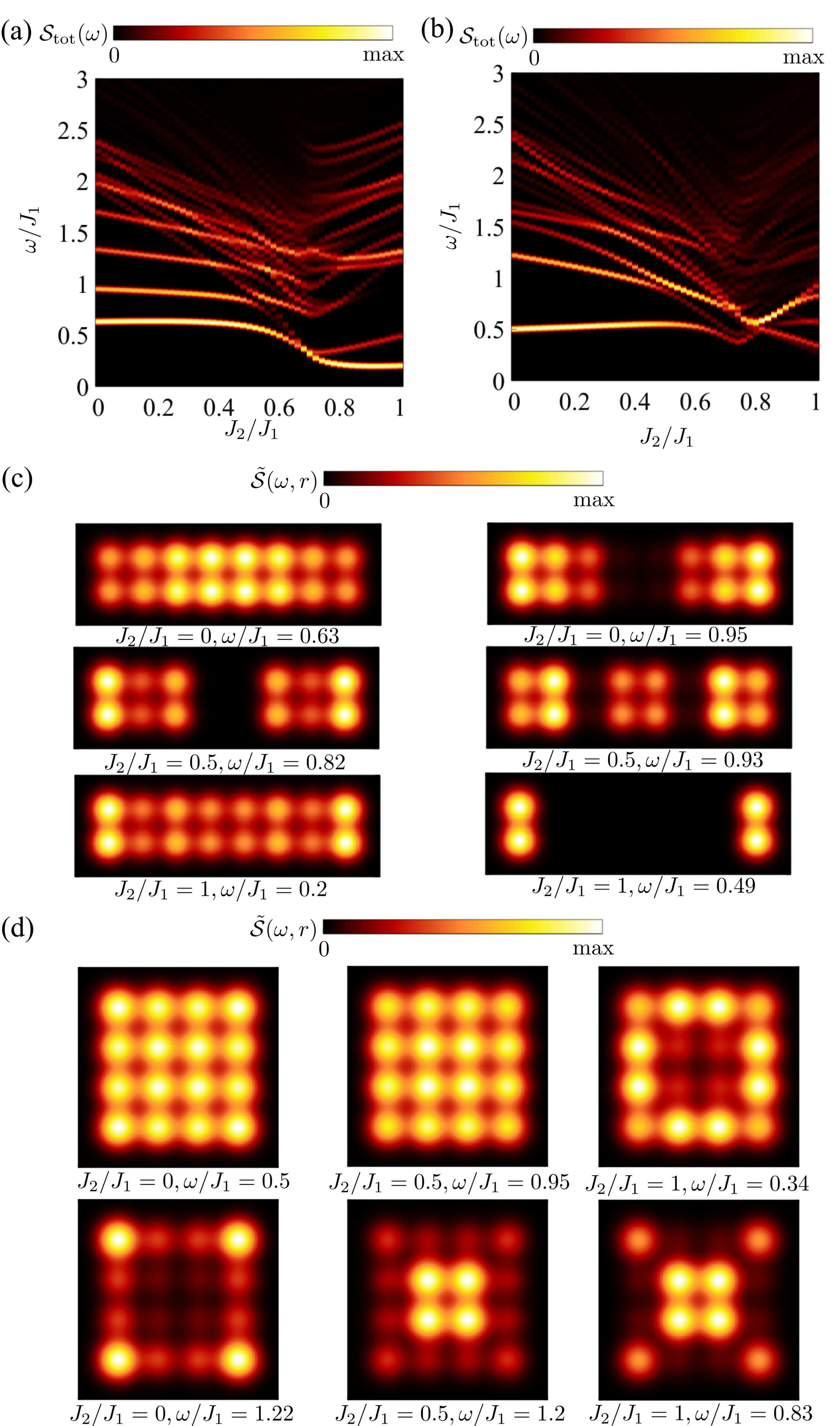}
\caption{Panels (a)/(b) shows the total spin spectral function in a $8\times 2$/$4\times 4$ ladder as a function of $J_2$. 
Panel (c) shows the local spectral function at different energies $\omega$
and $J_2$ for the $8\times 2$ ladder, and panel (d) shows the local spectral function at
different energies $\omega$ and $J_2$ for the $4\times 4$ ladder.
It is observed that both in the ladder and island the confined modes depend on the energy $\omega$
and the second neighbor coupling $J_2$.}
\label{fig:fig2}
\end{figure}

For the sake of concreteness, we will focus on the dynamical spin correlator for two different geometries a quasi-1D $8\times 2$ ladder and a quasi-2D $4\times 4$ dot. The spin systems are the most realistic ones to assemble from the already demonstrated structures, that achieved an engineered $2 \times 2$ cluster\cite{Yang2021}. Figure \ref{fig:fig2} shows the total dynamical correlator $S_{\text{tot}}(\omega)=\sum_{\mathbf{r}_i} S(\omega,\mathbf{r}_i)$ and the local correlator for different regimes of $J_2/J_1$ of the spin Hamiltonian. The total correlator (Figs \ref{fig:fig2}a and \ref{fig:fig2}c) shows a clear dependence on the  $J_2/J_1$, thus providing a hint on where in the phase diagram the system is. In Figs. \ref{fig:fig2}b and \ref{fig:fig2}d, we can observe the strong spatial dependence of the local spin correlator $\mathcal{S}(\omega,\mathbf{r}_i)$ for different $J_2/J_1$ values. This dramatic difference suggests that the change of the confined many-body spin excitations can provide a useful strategy to infer the form of the underlying spin Hamiltonian and determine the regime of the phase diagram in which the system is.

\begin{figure*}[t!]
\center
\includegraphics[width=\linewidth]{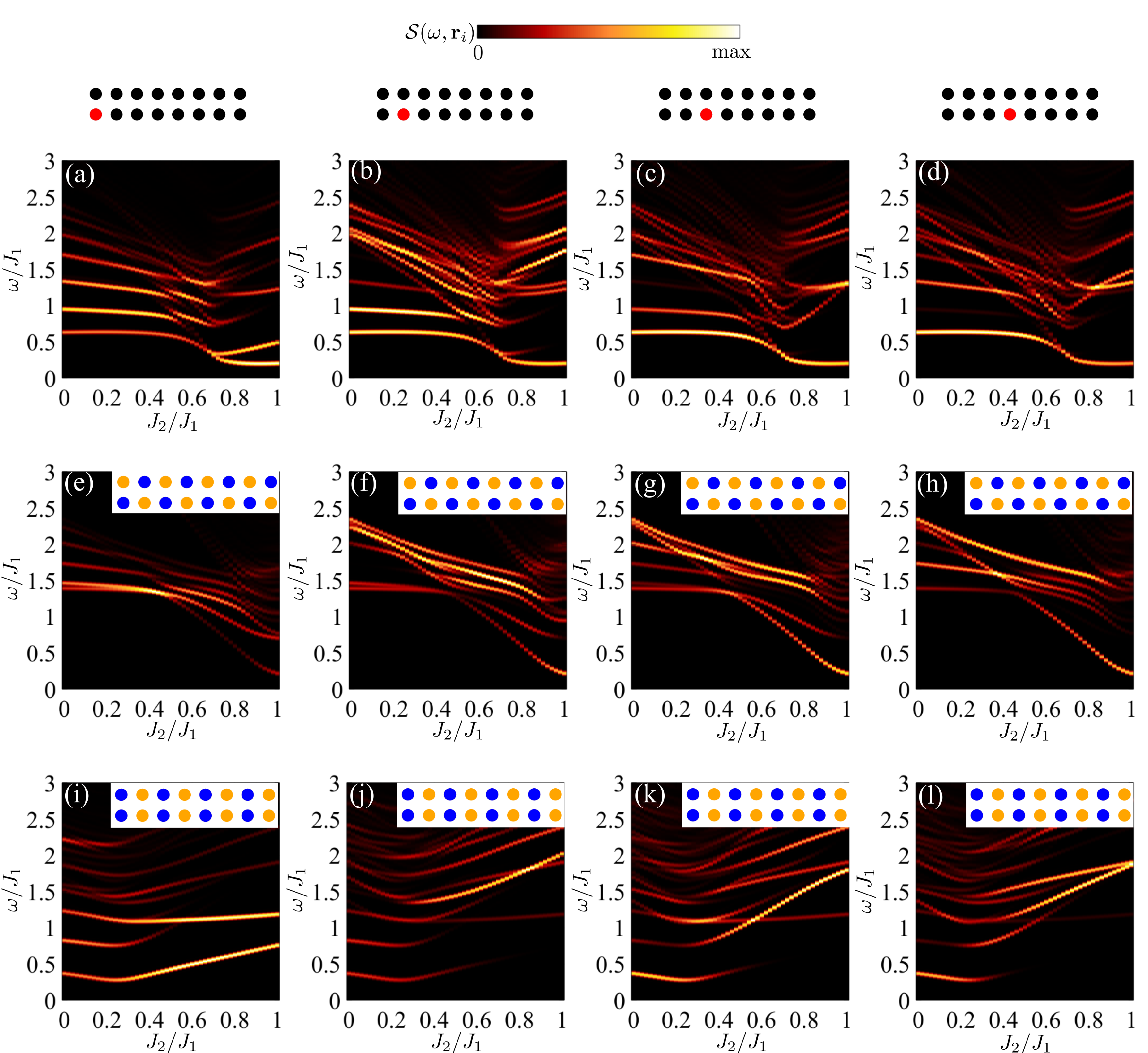}
\caption{Local spectral function at different energies $\omega$
and $J_2/J_1$ on different inequivalent sites for the $8\times 2$ ladder, with (a-d) no magnetic field, (e-h) $B_{AF}=0.5$ and (i-l) $B_{SAF}=0.5$. The insets in panels (e-l) show the local magnetic fields $B_i=\pm0.5 J_1$ on orange/blue site, respectively.}
\label{fig3}
\end{figure*}

\begin{figure}[t!]
\center
\includegraphics[width=\linewidth]{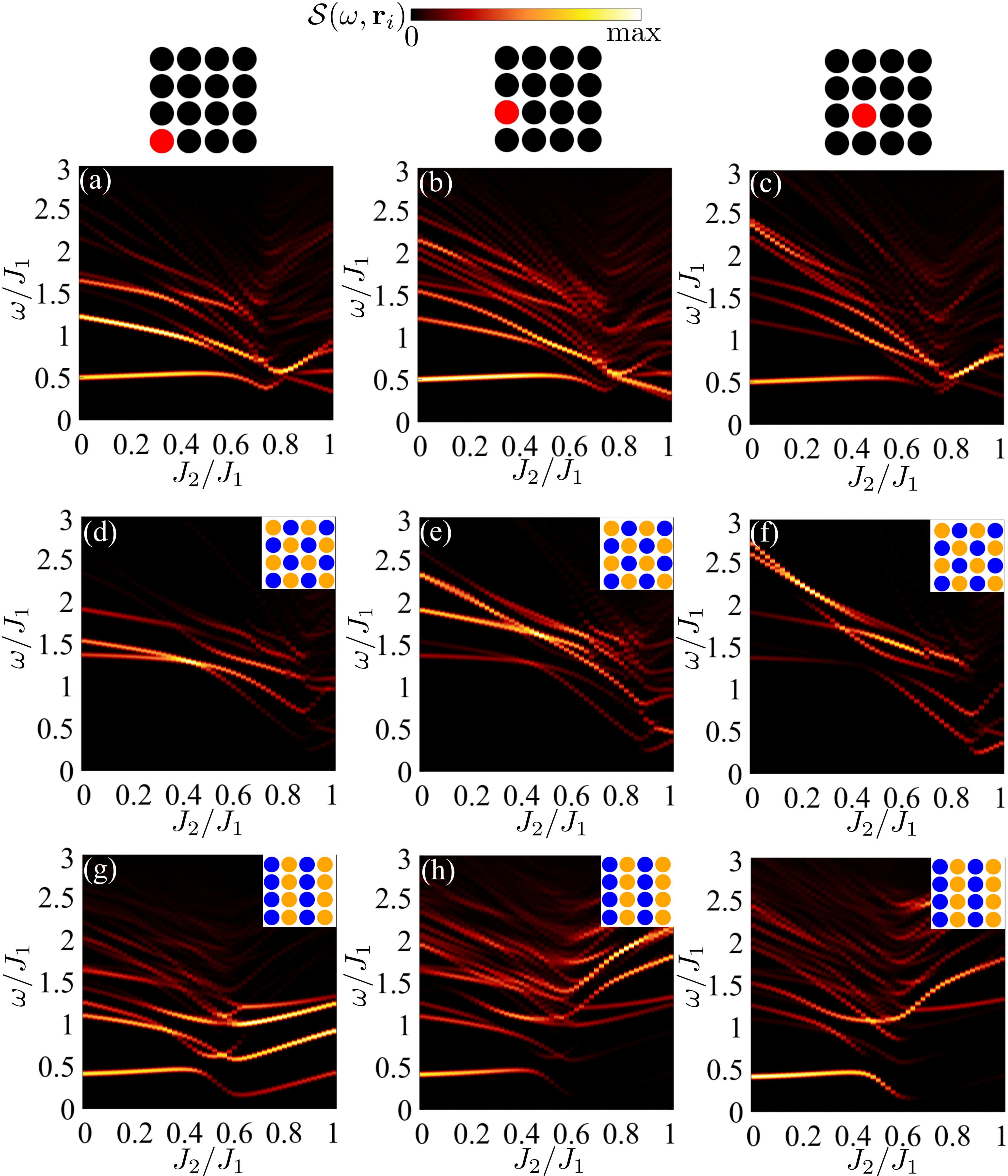}
\caption{ Local spectral function at different energies $\omega$
and $J_2$ on different sites for the $4\times 4$ island, with (a-d) no magnetic field, (e-h) $B_{AF}=0.5$, and (i-l) $B_{SAF}=0.5$. The insets in panels (d-i) show the local magnetic fields $B_i=\pm0.5 J_1$ on orange/blue site, respectively.}
\label{fig4}
\end{figure}

We can analyze in a more systematic way the evolution of the local spin correlator as a function $J_2/J_1$ for the different inequivalent sites of the islands that we are studying. Figures \ref{fig3}(a-d) show results for the $8\times 2$ ladder and Figs. \ref{fig4}(a-c) for the $4\times 4$ island. Moreover, to reveal the potential phase transitions more clearly, we consider adding a staggered anti-ferromagnetic field $B_{\text{AF}}$ (Figs. \ref{fig3}(e-h) and Figs. \ref{fig4}(d-f)) and a stripe anti-ferromagnetic field $B_{\text{SAF}}$ (Figs. \ref{fig3}(i-l) and Figs. \ref{fig4}(g-i)) to the Hamiltonian of eq. (\ref{eq1}) and study the changes that they produce on the local dynamical correlators.
We can start analyzing first the cases without external local Zeeman field (Figs. \ref{fig3}(a-d) and Figs. \ref{fig4}(a-c)). 
In all cases, the dynamical correlator shows several peaks at $J_2/J_1=0$. As $J_2/J_1$ increases, the peaks at higher energy move towards zero energy. Eventually, at around $J_2/J_1=0.8$ these peaks merge, then disperse for larger $J_2/J_1$.

Interestingly, the nanoscale nature of these magnets allows engineering local fields by depositing close to the spin-lattice atoms realizing a strong Ising spin, as is the case of Ho at MgO\cite{Natterer2017} or Dy at MgO\cite{Singha_2021}. These Ising spins allow the creation of local Zeeman fields in nearby spins and can be controlled individually, allowing to engineer atomically precise Zeeman profiles\cite{Singha_2021}. Atomically engineered Zeeman fields provide a strategy to probe the response of a quantum magnet to different Zeeman textures and, in particular, to distinguish between different states depending on their response. Using the previous idea, in the following, we include a local Zeeman term that will be induced by proximal Ising spins as

\begin{equation}
\mathcal{H}_Z = \sum_{\mathbf{r}_i}
\mathbf{B}_{\mathbf{r}_i} \cdot
\mathbf{S}_{\mathbf{r}_i}
\end{equation}

with $\mathbf{B}_{\mathbf{r}_i}$ the local Zeeman field created by proximal Ising spins. We note that while a Zeeman profile for the quasi-1D structures of Fig. \ref{Fig:schematic}c can be engineered in all the sites, Zeeman profiles in quasi-2D structures of Fig. \ref{Fig:schematic}b would be more challenging. As a reference, we will illustrate the impact of those profiles both for quasi-1D and quasi-2D, keeping in mind that the quasi-1D cases will be experimentally easier to implement. We will address to specific Zeeman profiles, stagger AF and collinear AF, that promote the two symmetry-broken states of Fig. \ref{Fig:schematic}d. The cases with the staggered anti-ferromagnetic field $B_{\text{AF}}$ (Figs. \ref{fig3}(e-h) and Figs. \ref{fig4}(d-f)) show a drastic change of the spectrum for  $0<J_2/J_1<0.6$: the lowest excitation at around $\omega=0.5 J_1$ vanishes. When $J_2/J_1>0.6$, the change is less significant: the levels are shifted, but no vanishing of strong peaks. This suggests that for $J_2/J_1<0.6$ the ground state has a large staggered AF component as it is sensitive to a staggered antiferromagnetic field, while for $J_2/J_1>0.6$ the ground state does not have large stagger antiferromagnetic component. Similarly, we can analyze the dynamical correlators in the presence of a stripe antiferromagnetic field $B_{\text{SAF}}$ (Figs. \ref{fig3}(i-l) and Figs. \ref{fig4}(g-i)): for $J_2/J_1<0.4$ the spectrum does not change that much, while for $J_2/J_1>0.4$ the lowest states vanish. This indicates a large stripe antiferromagnetic component in the ground state for $J_2>0.4$. This phenomenology reflects the behavior of the state in the thermodynamic limit\cite{doi:10.7566/JPSJ.84.024720}. For $J_2/J_1<0.4$, the ground state is almost staggered AF, while for $J_2/J_1>0.6$ the ground state is almost a stripe AF, and for $0.4<J_2/J_1<0.6$, the ground state is in a frustrated regime having competing staggered AF and stripe AF components.

The above analysis demonstrates three different phases in the $8\times 2$ ladder and the $4\times 4$ island as a function of $J_2/J_1$. Furthermore, it shows that the dependence of the spin excitations on a local local field changes dramatically depending on the value of $J_2/J_1$. This severe change of dynamical excitations with a local field directly reflects the nature of the different ground states as a function of $J_2/J_1$. In spite of having determined the connection between the local dynamical spin excitations and the different phases realized by the spin Hamiltonian, there are some limitations in order to accurately infer the form of the Hamiltonian, i.e. the precise $J_2/J_1$ ratio from the local spin dynamical correlator that one would measure in an experiment. In particular, we have been using a local magnetic field to analyze the evolution of the dynamical correlation with different antiferromagnetic external fields. While such local fields can be engineered in certain structures as noted above,\cite{Singha_2021}, from an experimental point of view, uniform magnetic fields are easier to control externally. As we will see below, even with a uniform magnetic field, the nature of the different ground states can be inferred from the local excitations by exploiting the dependence on confinement interference effects. The thickness of MgO is expected to have an impact on $J_1/J_2$ ratio due to the contribution to the exchange mediated by the underlying silver substrate in experiments. Specifically, thin MgO substrates will give rise to a substantial enhancement of $J_2/J_1$. In contrast, for large thicknesses of MgO, $J_1$ is expected to be dominating due to exchange being dominated by super-exchange through MgO. The impact of silver in the magnetic properties of Ti and other 3d atoms has been addressed previously\cite{PhysRevLett.119.227206,Willke2018,Shehada2022,Tosoni2022}, demonstrating the notable role of the MgO substrate in modifying the effective spin Hamiltonian. The theoretical analysis above has been able to approximately establish the different phase transition from the evolution of the dynamical correlators. However, it would be better to provide a way to directly get the precise $J_2/J_1$ ratio from the correlator that would be directly measured in experiments. In the following section, we will establish a protocol to systematically infer the specific form of the Hamiltonian from the dynamical correlator.

\section{Hamiltonian learning from confined quantum spin excitations}

In the previous section, we observed that the spatially resolved dynamical excitations show a distinct dependence with the Hamiltonian parameters. While extracting the parameters from the data is a non-trivial task. Machine learning methods (ML) allow to process data without the need for explicitly programmed and task-depended algorithms. The algorithm learns to solve the task and creates a model purely from data, thus overcoming the issues that we have described.

\begin{figure}[t!]
\center
\includegraphics[width=\linewidth]{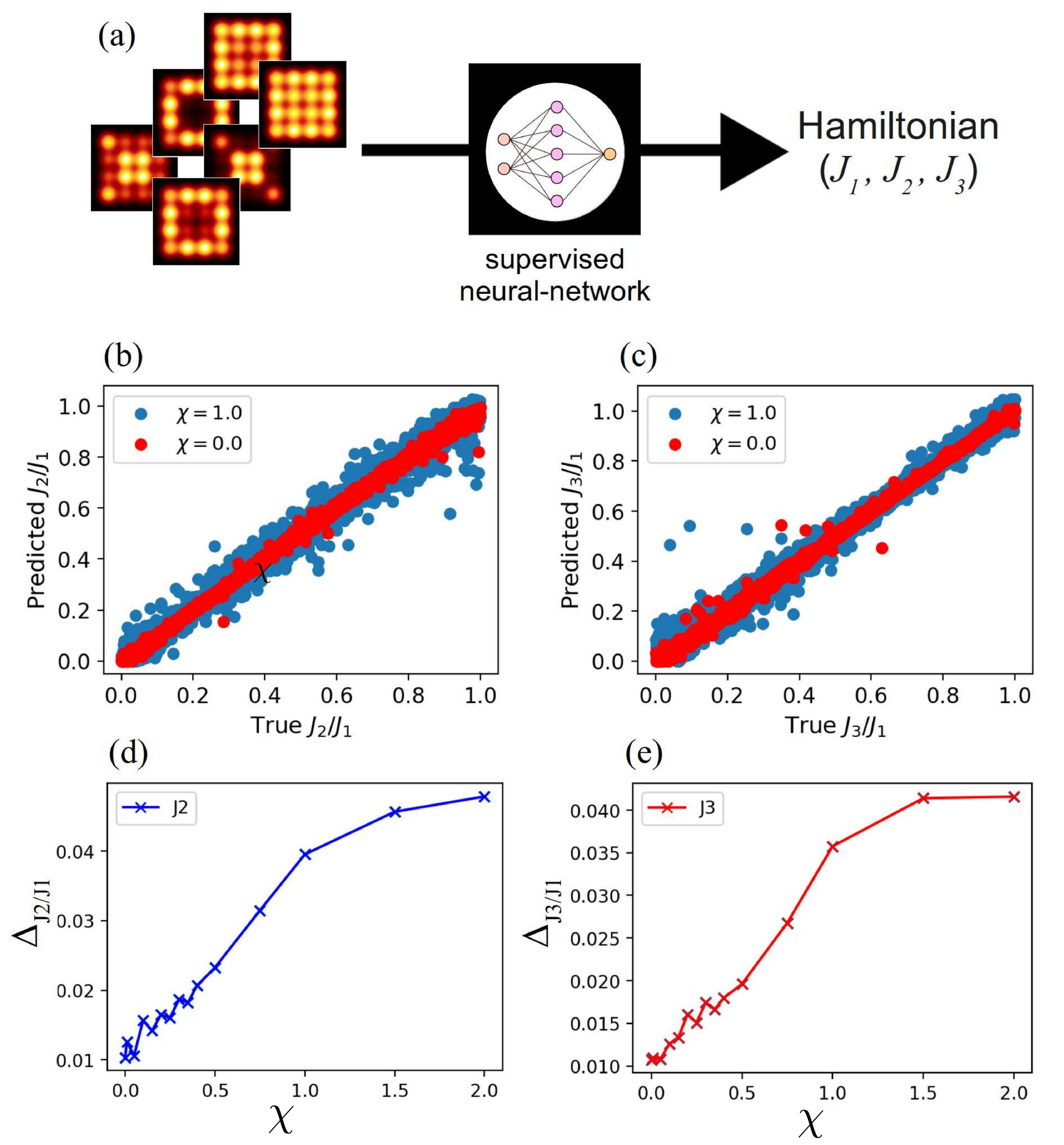}
\caption{(a) Schematic workflow of the NN, taking dynamical correlators as input and returning the underlying Hamiltonian parameter ($J_1, J_2, J_3$). Panels (b)-(e) show the results for the $2\times 8$ lattice including the predictions vs. true values of $J_2$ (b) and $J_3$ (c) with ($\chi=1$) and without noise. Panels (d) and (e) show the corresponding mean error (ME) $\Delta_{J_2/J_1}$ and
$\Delta_{J_3/J_1}$ for noise up to $\chi=2$ for $J_2$ and $J_3$. }
\label{fig6}
\end{figure}

In the following we will use supervised learning with Neural Networks (NNs). NNs are used to perform a regression task, which assumes a relationship between the inputs and outputs of the model. Here, the inputs are (local) dynamical correlators (given by Eq.~\ref{eq2}) and the outputs the corresponding exchange parameters of the Hamiltonian. A sketch of the workflow of the NN is shown in Fig. \ref{fig6}a. The NN acts as a universal function approximator, finding a function $f$, mapping from the inputs $X$ to the outputs $y$. By optimization of the NN parameter we obtain $y=f(X)$. For the optimization task we use supervised learning and a gradient decent algorithm, i.e. teaching the NN by showing examples of pairs of inputs and outputs to minimize the loss function and update the NN parameter (weights). The input of the NN contains the map of dynamical correlators at the difference frequencies. By exploiting the symmetry of the lattices, we reduce the whole map to 3 (4) independent dynamical correlator of the $4\times 4$ ($8\times 2$) system in combination with an external uniform magnetic field which is the one easier to implement in experiments. This results in a total dimension of 601 (801). The NN architecture consists of 3 hidden layers of dimensions 200, 200, and 100 and the output dimension of 2. To demonstrate our algorithm, we demonstrate our algorithm with a more complex Heisenberg Hamiltonian, incorporating both first, second and third neigh exchange $(J_2/J_1, J_3/J_1)$ of the form

\begin{multline}
	\label{eq3}
H=J_1\sum_{\langle {\mathbf{r}_i},\mathbf{r}_j\rangle}\mathbf{S}_{\mathbf{r}_i} \cdot \mathbf{S}_{\mathbf{r}_j}
+J_2\sum_{\langle\langle {\mathbf{r}_i},{\mathbf{r}_j}\rangle\rangle}\mathbf{S}_{\mathbf{r}_i} \cdot \mathbf{S}_{\mathbf{r}_j} \\
+J_3\sum_{\langle\langle\langle {\mathbf{r}_i},{\mathbf{r}_j}\rangle\rangle\rangle}\mathbf{S}_{\mathbf{r}_i} \cdot \mathbf{S}_{\mathbf{r}_j}
+ \mathbf{B} \cdot \sum_{{\mathbf{r}_i}} \mathbf{S}_{\mathbf{r}_i} \, ,
\end{multline}

with $\mathbf{B}=(B_x,0,0)$, and where we added third-neighbour interactions ($J_3$) as well as an external magnetic field with strength $B_x$. We set $J_1=1$ as the energy scale and vary the parameters $(J_2,J_3,B_x)$ in the interval $I=[0,1]$ to cover a wide range of the parameter space of the Hamiltonian.
The magnetic field $B_x$ is measured in units of $g\mu_B/J_1$, with $g=1$ the Land\'e g-factor and $\mu_B$ he Bohr magneton.
We use the Hamiltonian parameters $(J_2,J_3)$ as labels in the dataset to train the NN. The trained model is then able to predict the underlying exchange Hamiltonian parameters $J_2$ and $J_3$, by using the dynamical correlators and magnetic field strength as inputs. 
We train the NN for 100 epochs with a batch size of 10 to obtain the results shown in this work. 
We will use this procedure to address the $2\times 8$ and $4 \times 4$ lattices presented in the previous section. The general ML workflow is shown in Fig.~\ref{fig6}(a). The NN takes as input dynamical correlators with corresponding magnetic field $B_x$ to predict the corresponding $J_2$ and $J_3$ interactions. 

\begin{figure}[t!]
\center
\includegraphics[width=\linewidth]{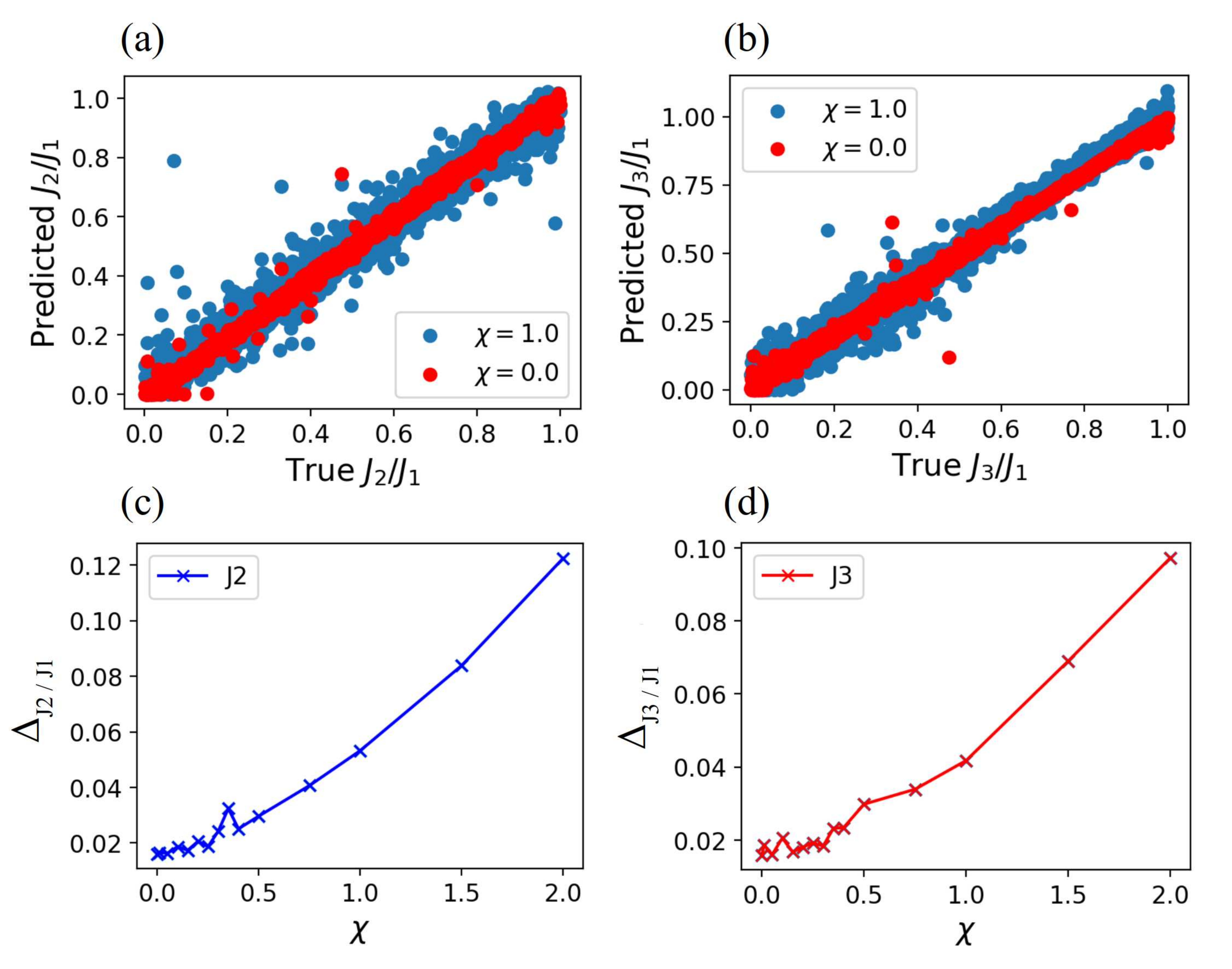}
\caption{ Panels (a)-(d) show the results for the $4\times 4$ lattice including the predictions vs. true values of $J_2$ (a) and $J_3$ (b) with noise ($\chi=1$) and without noise. Panels (c) and (d) show the corresponding mean error of $J_2$ (c) and $J_3$ (d) for added noise up to $\chi=2$. }
\label{fig7}
\end{figure}

In order to demonstrate that NN Hamiltonian inference is a robust methodology to be applied to experimental data, we have introduced noise in the dynamical correlators used to infer $J_2$ and $J_3$ interactions.
Noise is included as a random frequency-dependent renormalization of the dynamical correlators defined as

\begin{equation}
    \mathcal{S}(\omega,\mathbf{r}_i)_{noisy} = \mathcal{S}(\omega,\mathbf{r}_i) \cdot |1 + \zeta(\omega)| \, .
\end{equation}

The noise $\zeta(\omega)$ is defined as a uniform distribution defined in the interval $\zeta(\omega) \in [-\chi, \chi]$, randomly sampled for each Hamiltonian, and different for every discrete frequency of the dynamical correlators.

Figure~\ref{fig6}(b)-(e) shows the results for the $2\times 8$ lattice, starting with the predictions for $J_2$ and $J_3$ (in Fig.~\ref{fig6}(b) and Fig.~\ref{fig6}(c)) in the present and absence of noise. 
The predictions are plotted against the true values and in the ideal case, all data points are exactly lying on the diagonal line [of f(x)=x]. Without the presence of noise (red) the NN predicts both values with very high precision, with a mean error (ME) of $\Delta_{J_2/J_1}$ = 0.0103 and $\Delta_{J_3/J_1}$ = 0.0107. However, even in the presence of very high noise levels ($\chi=1$), the NN is able to make predictions with errors of $\Delta_{J_2/J_1}$ = 0.040 and $\Delta_{J_3/J_1}$ = 0.036. This shows the high resilience to noise of our approach, which is an important factor when considering experimental data. 
The resilience to noise is highlighted even more in Figs. \ref{fig6}(d) and \ref{fig6}(e), where the MSE is shown for different noise levels for $J_2$ (d) and $J_3$ (e). As seen in both figures, the ME stays very low up to noise levels of $\chi=0.5$ to the dynamical correlators when it slowly starts increasing. Even for noise levels of $\chi=1$ and above the errors remain relatively small. The NN is still able to make good predictions for noise levels of $\chi=1$ and above. For the following predictions, we will use added noise with amplitude $\chi=0.02$ since it can be beneficial for the training of the NN in order to avoid over-fitting.

\begin{figure*}[t!]
\center
\includegraphics[width=0.8\linewidth]{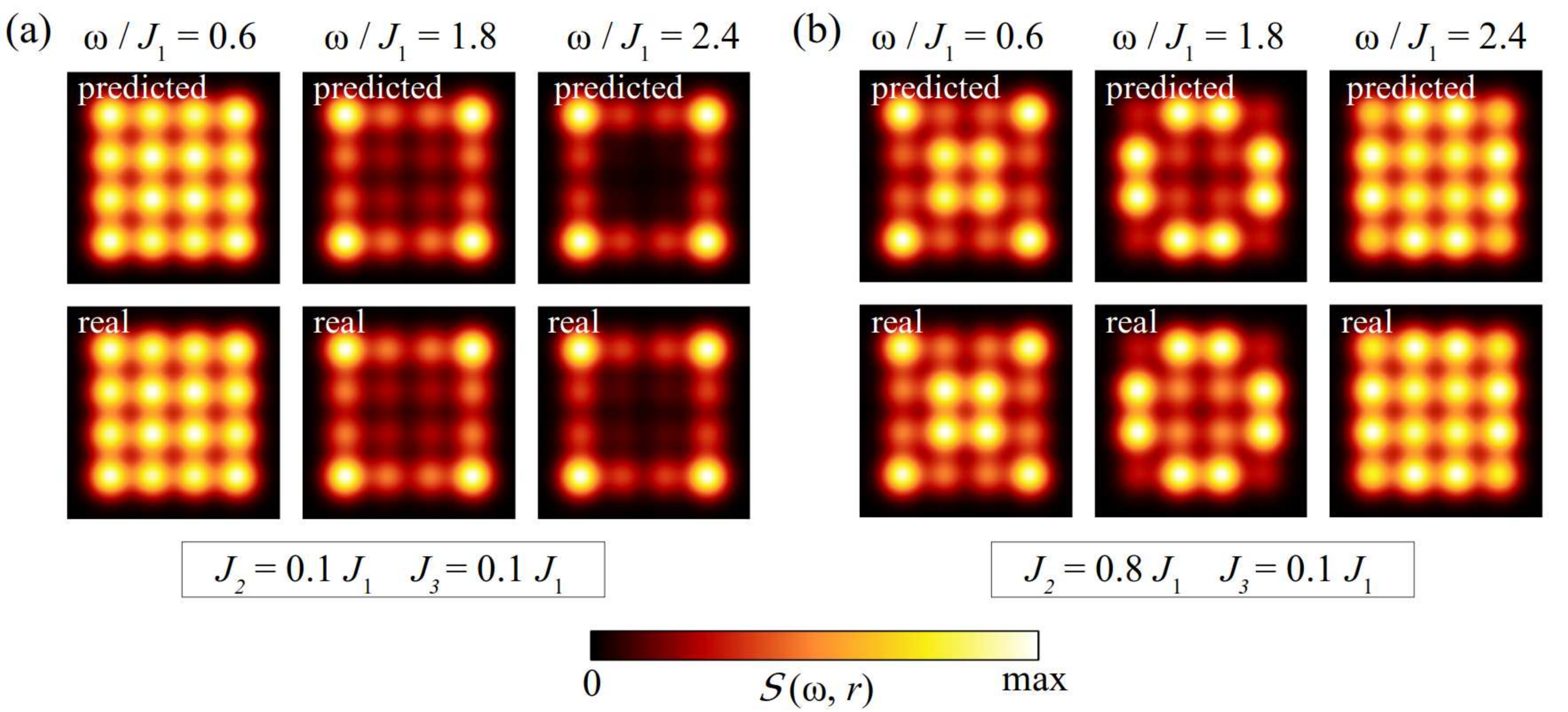}
\caption{ Local spectral function at 3 different energies $\omega$ for the $4\times 4$ system. Shown are the predictions of the NN (top) in comparison with numerical computations (bottom) for the Hamiltonians $J_2=0.1 J_1$, $J_3=0.1 J_1$ ($B_x = 0$) (a) and $J_2=0.8 J_1$, $J_3=0.1 J_1$ ($B_x = 0$) (b).}
\label{fig8}
\end{figure*}

\begin{figure}[t!]
\center
\includegraphics[width=\linewidth]{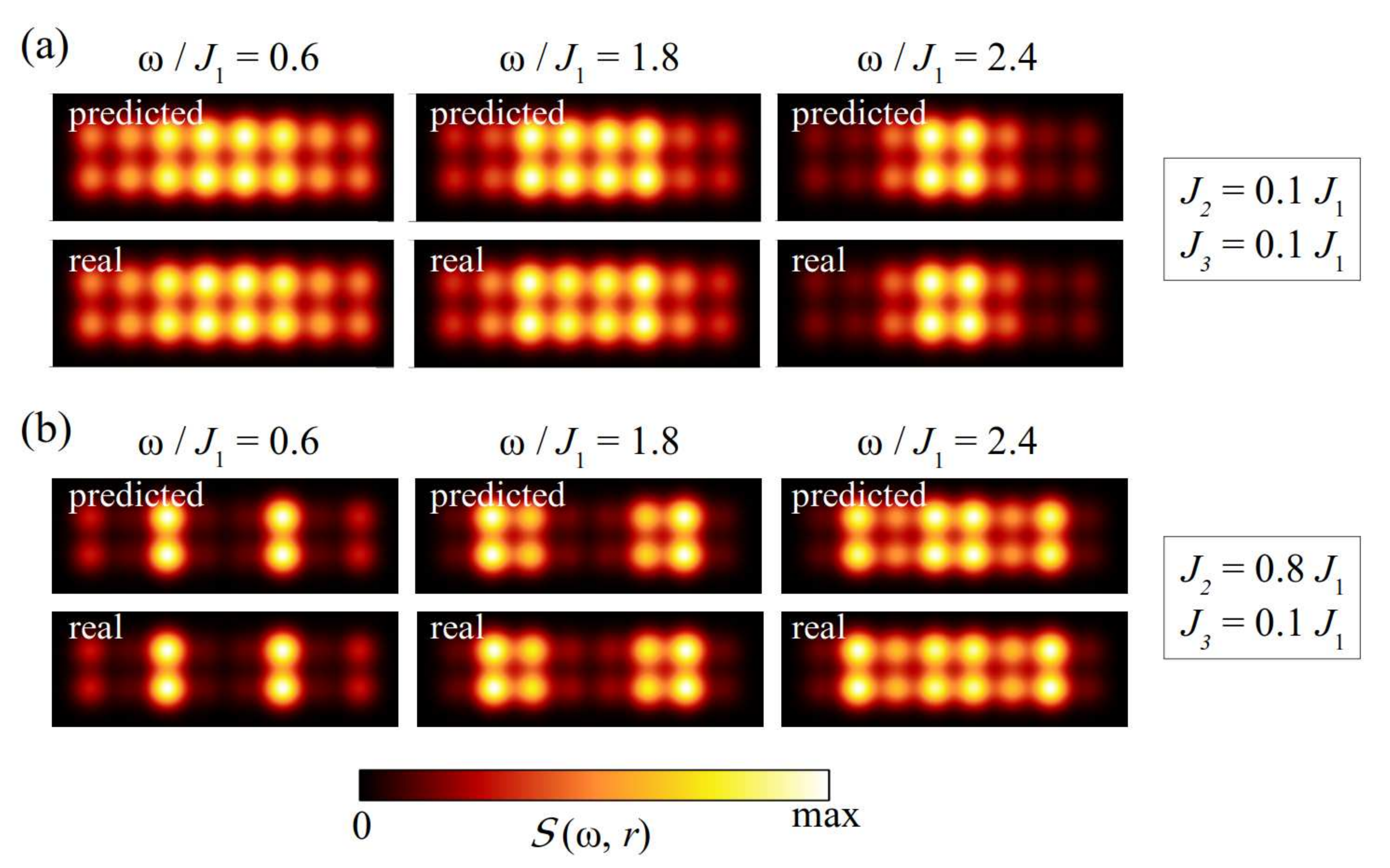}
\caption{ Local spectral function at 3 different energies $\omega$ for the $2\times 8$ ladder. Shown are the predictions of the NN (top) in comparison with numerical computations (bottom) for the Hamiltonians $J_2=0.1 J_1$, $J_3=0.1 J_1$ ($B_x = 0$) (a) and $J_2=0.8 J_1$, $J_3=0.1 J_1$ ($B_x = 0$) (b).}
\label{fig9}
\end{figure}

The same analysis is done for the $4\times 4$ lattice in Fig. \ref{fig7}. In this case, we trained the network with dynamical correlators of the $4\times 4$ system. The results are almost identical to the $2\times 8$ lattice shown in Fig. \ref{fig6}. Figures \ref{fig7}(a) and \ref{fig7}(b) show the predictions vs. true Hamiltonian parameter for $J_2$ and $J_3$ respectively which are of very high precision for no noise and even up to noise amplitudes of $\chi=1$ the NN predicts the parameters accurately. The ME for both parameters without noise is $\Delta_{J_2/J_1}$ = 0.016  and $\Delta_{J_3/J_1}$ = 0.016 and with added noise of $\chi=1$, the NN is again able to make predictions with low errors of $\Delta_{J_2/J_1}$ = 0.053 and $\Delta_{J_3/J_1}$ = 0.042.
We also see a similar behavior for the ME for different noise levels in Fig. \ref{fig7}(c) and Fig. \ref{fig7}(d). Also for this system the ME starts increasing from  noise amplitudes of $\chi=1$ but does not saturate at $\chi=1.5$ as it was the case for the $2\times 8$ lattice . Nonetheless, the NN is very resilient to noise and able to make good predictions.
Figure \ref{fig8} shows the predictions of the local spin correlation function made by the NN in comparison with numerical calculations for the $4\times 4$ system in two different regions of the Hamiltonian. It is shown, that the NN performs well for the AFM phase Fig. \ref{fig8}(a) as well as the VBS phase Fig. \ref{fig8}(b). The predictions are almost indistinguishable from the numerical calculations, including all significant features, and only show very minor deviations.
The same results are obtained for the $2\times 8$ ladder presented in Fig. \ref{fig9} for the same Hamiltonian parameter, where the NN performs equally well to predict the local spectra function for 3 different energies. The high precision of the NN predictions is related to the high accuracy in the predictions of the Hamiltonian parameter and the small mean error discussed before.

Finally, we discuss some additional considerations regarding the experimental parameter extraction in generic spin lattices beyond Ti in MgO.  While for Ti in MgO, the spin excitations are dominated by the isotropic Heisenberg coupling model, in generic experimental realizations, there can be additional contributions to the Hamiltonian. In particular, contribution to the spin coupling stemming dipolar interaction can appear\cite{Choi2017}. This contribution can be accurately computed from the geometry of the island. As a result, such a term can be explicitly included in the Hamiltonian, and its prefactor does not have to be inferred. In addition, small anisotropic exchange contributions coming from spin-orbit coupling can appear in real-spin models. While such a contribution is small for Ti lattices, lattices made of heavier 4d or 5d atoms can display stronger anisotropic exchange stemming from spin-orbit coupling\cite{Natterer2017}. In those instances, the anisotropic term should be included in the Hamiltonian, with a parameter that has to be inferred in the supervised learning. Furthermore, in the case of lattices with $S>1/2$, such as Fe in MgO\cite{Andreas2015}, local spin anisotropy terms would have to be included in the Hamiltonian and inferred in the learning. Finally, in the case the Hamiltonian inference is made in the presence of a large external magnetic field, anisotropic g-factor may appear\cite{PhysRevResearch.1.033185} and could become an additional parameter to be extracted.
In the appendices, we show that the NN formalism can be extended to infer both the anisotropy strength and the g-factor. This demonstrates the potential of machine learning algorithms to infer generalized Hamiltonians with multiple parameters.

Finally, it is worth emphasizing the clear advantage of our algorithm with respect to conventional fitting procedures. Our machine learning algorithm requires evaluating the spin spectral function only to train the algorithm initially, a task that can be systematically parallelized. This means that, once the algorithm is trained, parameters can be obtained instantly, extracted just by providing the measured spectral function. In contrast,
conventional fitting algorithms require iteratively evaluating the spin spectral function of the model for different parameters up to several hundred times, a task that can become time consuming. This evaluation cannot be done in parallel, as each parameter chosen depends on the quality of the fitting for the previous one. This implies that our methodology provides a nearly instantaneous Hamiltonian extraction in comparison with the standard fitting method. This massive speed-up is specially relevant for using our algorithm with automated impurity assembly recently demonstrated\cite{Chen2022ML}, as our methodology would allow to estimate on the fly the Hamiltonian of each realized atomic arrangement.

\section{Conclusion}
To summarize, we have shown that spatially and frequency-resolved spin excitations in finite-size spin models allow inferring the underlying long-range frustrated Heisenberg Hamiltonian.  Our methodology exploits the finite-size effects of a quantum-spin island, demonstrating that confinement in the many-body spin modes provides a strategy to extract microscopic couplings.  From the experimental point of view, our results show that spatial and frequency resolution of electrically driven spin resonance with scanning tunneling microscopy allows extracting the nature of complex Hamiltonians from confinement effects in finite spin systems.  In particular, by focusing on the $S=1/2$ Heisenberg model as realized by Ti in MgO, we showed that the spatial distribution of spin excitations strongly depends on the underlying exchange couplings of the model.  This strong dependence displayed in finite systems can be rationalized by the different nature of the ground state in the thermodynamic limit, which impacts the spin excitations even in small, confined islands.  We demonstrated that such finite-size effects allow the development of a supervised-learning methodology for extracting the parameters of the Hamiltonian from the full frequency and spatially resolved spin excitations.  We showed that this methodology is robust to noise in the dynamical spin excitations, establishing an experimentally realistic strategy for Hamiltonian inference using real data from paramagnetic resonance measurements with scanning tunnel microscopy.  Our methodology puts forward confined excitations in frustrated magnets as a powerful strategy to understand the build-up and nature of frustrated quantum spin many-body models.

\section*{Acknowledgements}

We acknowledge the computational resources provided by the Aalto Science-IT project,
the financial support from the Academy of Finland Projects No. 331342, No. 336243 and No 349696,
and the Jane and Aatos Erkko Foundation. We thank K. Yang, T. Kurten, P. Liljeroth, S. Kezilebieke and R. Drost for useful discussions.

\appendix

\section{g-factor predictions}
\begin{figure}[t!]
\center
\includegraphics[width=\linewidth]{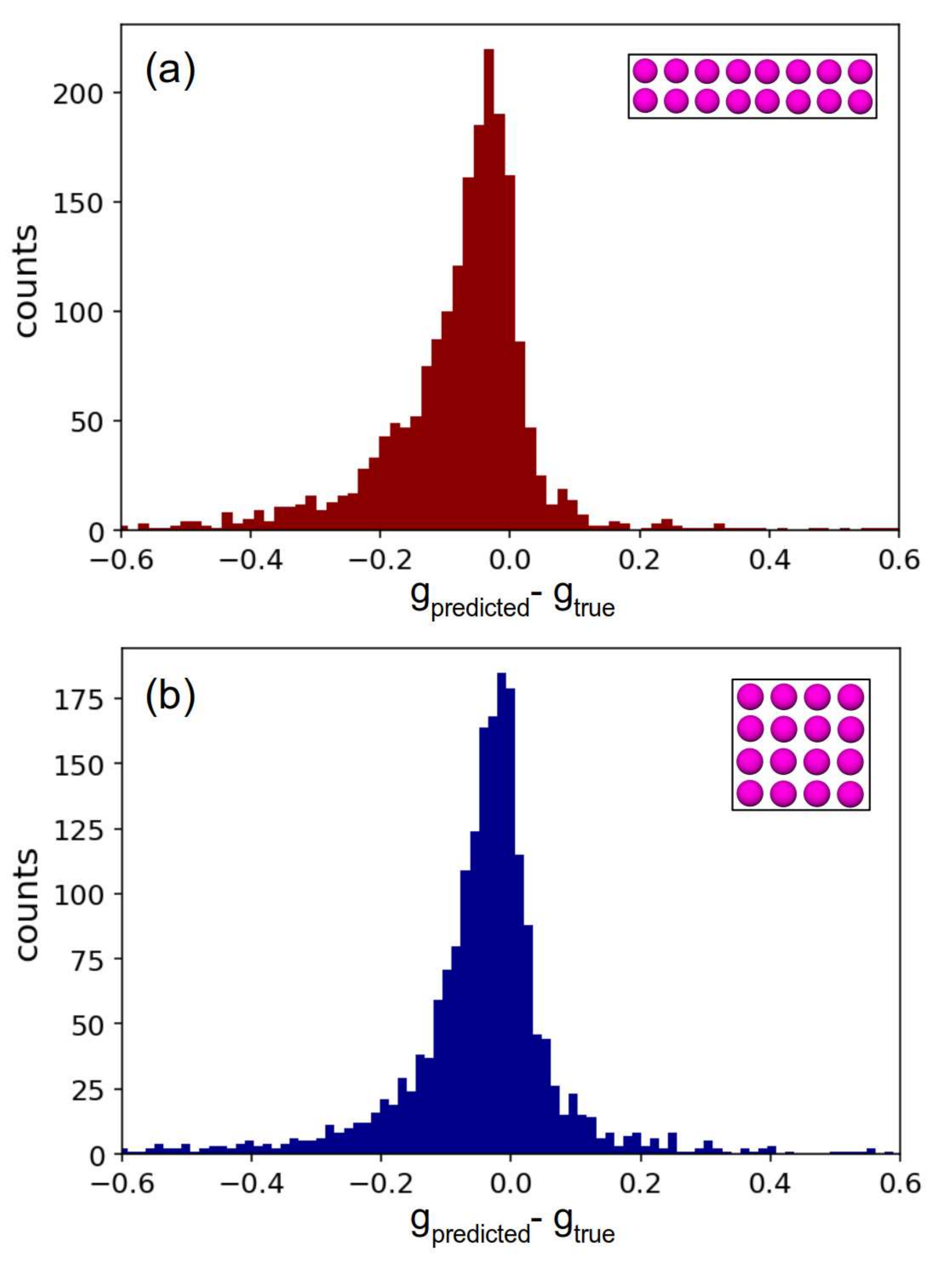}
\caption{ Histogram of g-factor predictions for the $2\times8$ lattice (a) and $4\times4$ lattice (b). The number of counts is plotted against the difference of the predictions and true value. The mean absolute error for (a) is $\Delta=0.101$ and for (b) $\Delta=0.092$. }
\label{gfactor}
\end{figure}

While in the main manuscript we focused on predicting exchange constants, additional terms may appear in the Hamiltonian.
Specifically, spin-orbit coupling effects give rise to a renormalized g-factor affecting the Zeeman term. We now show that our algorithm
would be capable of predicting this additional term.
Figure~\ref{gfactor} shows the predictions for the g-factor for the $2\times8$ lattice (a) and $4\times4$ lattice (b). We included the g-factor in the Hamiltonian as 
\begin{multline}
H=J_1\sum_{\langle {\mathbf{r}_i},\mathbf{r}_j\rangle}\mathbf{S}_{\mathbf{r}_i} \cdot \mathbf{S}_{\mathbf{r}_j}
+J_2\sum_{\langle\langle {\mathbf{r}_i},{\mathbf{r}_j}\rangle\rangle}\mathbf{S}_{\mathbf{r}_i} \cdot \mathbf{S}_{\mathbf{r}_j} \\
+J_3\sum_{\langle\langle\langle {\mathbf{r}_i},{\mathbf{r}_j}\rangle\rangle\rangle}\mathbf{S}_{\mathbf{r}_i} \cdot \mathbf{S}_{\mathbf{r}_j}
+ g \mathbf{B} \cdot \sum_{{\mathbf{r}_i}} \mathbf{S}_{\mathbf{r}_i} \, ,
\end{multline}
where the Zeeman term is renormalized by the g-factor $g$. The NN is trained to predict the g-factor, chosen in the interval [0.2, 2.0]. These results show that our ML algorithm is capable of predicting the g-factor in addition to $J_2$ and $J_3$ with good accuracy.

\section{Anisotropic exchange predictions }

\begin{figure}[t!]
\center
\includegraphics[width=\linewidth]{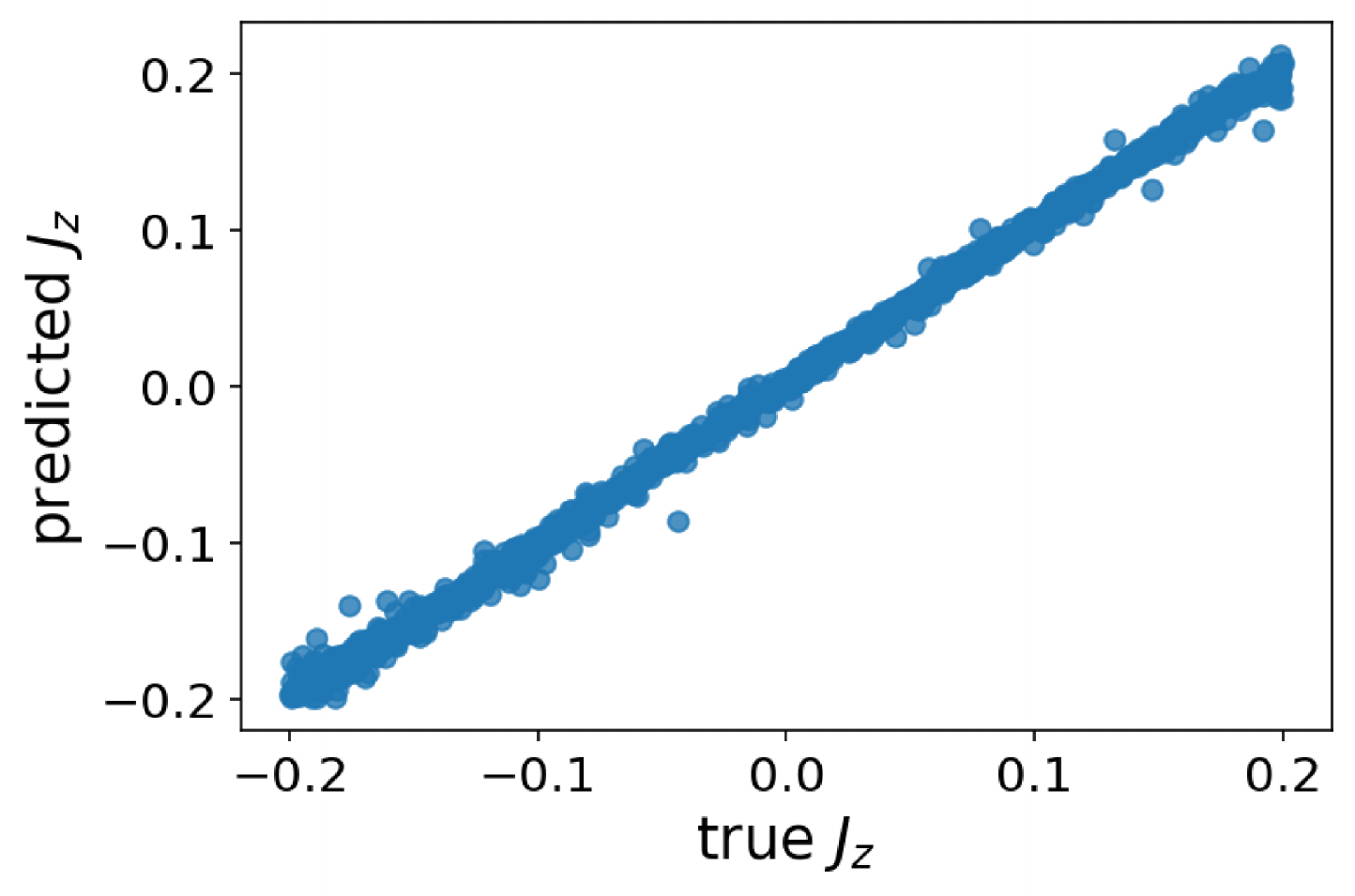}
\caption{ Predictions of the anisotropy strength $J_z$ vs. true values for the testset data $4\times4$ lattice. The mean absolute error is $\Delta=0.014$. }
\label{Jz}
\end{figure}

An additional term that emerges in spin Hamiltonian due to spin-orbit coupling is anisotropic exchange terms. In the following we show that our algorithm can be easily extended to extract anisotropic exchange interactions.
Figure~\ref{Jz} shows the predictions for the anisotropy first neighbor exchange correction $J_z$ for the $4\times4$ lattice. We included the anisotropy term in the Hamiltonian as 

\begin{multline}
	\label{eq3}
H=J_1\sum_{\langle {\mathbf{r}_i},\mathbf{r}_j\rangle}\mathbf{S}_{\mathbf{r}_i} \cdot \mathbf{S}_{\mathbf{r}_j}
+J_2\sum_{\langle\langle {\mathbf{r}_i},{\mathbf{r}_j}\rangle\rangle}\mathbf{S}_{\mathbf{r}_i} \cdot \mathbf{S}_{\mathbf{r}_j} \\
+J_3\sum_{\langle\langle\langle {\mathbf{r}_i},{\mathbf{r}_j}\rangle\rangle\rangle}\mathbf{S}_{\mathbf{r}_i} \cdot \mathbf{S}_{\mathbf{r}_j}
+ \mathbf{B} \cdot \sum_{{\mathbf{r}_i}} \mathbf{S}_{\mathbf{r}_i} \\
+ J_z \sum _{\langle {\mathbf{r}_i},\mathbf{r}_j\rangle} S^Z_{\mathbf{r}_i} \cdot S^Z_{\mathbf{r}_j} \, ,
\end{multline}

The NN is trained to predict the anisotropy strength $J_z$, chosen in the interval [-0.2, 0.2], by using the dynamical correlator $D(\omega)$ as input. The mean error is $\Delta=0.014$.
These results show our ML algorithm can be extended to account for anisotropic exchange terms in a spin Hamiltonian.

\bibliography{references.bib}

\end{document}